\documentclass[twocolumn]{aastex631}

\usepackage{comment} 
\usepackage{lipsum} 
\usepackage{graphicx} 
\usepackage{pgffor}
\usepackage{amsmath}
\usepackage{amsfonts}
\usepackage{subfigure}
\usepackage{makecell}
\usepackage{physics}

\usepackage{xcolor}

\usepackage{appendix}

\sloppypar
\newcommand{\params}{\vb*\theta}
\newcommand{\Sfeat}[2]{\vb{x}_{#1}^{#2}}
\newcommand{\Gfeat}[1]{\vb*X^{#1}}
\newcommand{\avlike}[2]{\overline{\rho}(#1|#2)}

\newcommand{\om}{$\Omega_{\rm m}$}

\shorttitle{The effects of parameters on galaxy properties in CAMELS}
\shortauthors{Contardo et al.}
\graphicspath{{./}{Figs/}}

\begin{document}


\title{On the effects of parameters on galaxy properties in CAMELS and the predictability of $\Omega_{\rm m}$}


\author[0000-0002-3011-4784]{Gabriella Contardo}
\email{gcontardo@ung.si}
\affiliation{Center for Astrophysics and Cosmology, University of Nova Gorica, Ajdovščina, Slovenia}
\affiliation{Theoretical and Scientific Data Science, Scuola Internazionale Superiore di Studi Avanzati (SISSA), Trieste, Italy}

\author[0000-0002-3415-0707]{Roberto Trotta}
\affiliation{Theoretical and Scientific Data Science, Scuola Internazionale Superiore di Studi Avanzati (SISSA), Trieste, Italy}
\affiliation{ICSC - Centro Nazionale di Ricerca in High Performance Computing, Big Data e Quantum Computing, Via Magnanelli 2, Bologna, Italy}
\affiliation{Astrophysics Group, Physics Department, Blackett Lab, Imperial College London, Prince Consort Road, London SW7 2AZ, UK}
\affiliation{INAF -- Osservatorio Astronomico di Trieste, Via G. B. Tiepolo 11, I-34143 Trieste, Italy}

\author[0000-0002-3375-1850]{Serafina Di Gioia}
\affiliation{Abdus Salam International Center for Theoretical Physics (ICTP), Trieste, Italy}
\affiliation{Theoretical and Scientific Data Science, Scuola Internazionale Superiore di Studi Avanzati (SISSA), Trieste, Italy}
\author[0000-0003-2866-9403]{David W. Hogg}
\affiliation{Center for Cosmology and Particle Physics, Department of Physics, New~York~University, New~York NY, USA}
\affiliation{Max-Planck-Institut f\"ur Astronomie, Heidelberg, Germany}
\affiliation{Center for Computational Astrophysics, Flatiron Institute, New~York NY, USA}

\author[0000-0002-4816-0455]{Francisco Villaescusa-Navarro}
\affiliation{Center for Computational Astrophysics, Flatiron Institute, New~York NY, USA}
\affiliation{Department of Astrophysical Sciences, Peyton Hall, Princeton University, NJ 08544-0010, USA}

\correspondingauthor{Gabriella Contardo}

\begin{abstract}
    Recent analyses of cosmological hydrodynamic simulations from \texttt{CAMELS} have shown that machine learning models can predict the parameter describing the total matter content of the universe, $\Omega_{\rm m}$, from the features of a single galaxy. We investigate the statistical properties of two of these simulation suites, \texttt{IllustrisTNG} and \texttt{ASTRID}, confirming that $\Omega_{\rm m}$ induces a strong displacement on the distribution of galaxy features. We also observe that most other parameters have little to no effect on the distribution, except for the stellar-feedback parameter $A_{SN1}$, which introduces some near-degeneracies that can be broken with specific features. These two properties explain the predictability of \om{}. We use Optimal Transport to further measure the effect of parameters on the distribution of galaxy properties, which is found to be consistent with physical expectations. However, we observe discrepancies between the two simulation suites, both in the effect of $\Omega_{\rm m}$ on the galaxy properties and in the distributions themselves at identical parameter values. Thus, although $\Omega_{\rm m}$'s signature can be easily detected within a given simulation suite using just a single galaxy, applying this result to real observational data may prove significantly more challenging.

\end{abstract}
%

 \section{Introduction}

A series of recent works \citep{onegalcosmo, onegalastrid, voidonegal,  Chawak2024ApJ} have investigated the possibility of predicting cosmological and astrophysical parameters from the astronomical features of a single (or few) galaxies, using machine learning (ML) and the \texttt{CAMELS} simulations, which vary a set of cosmological parameters (e.g. \om, $\sigma_8$) and sets of astrophysical parameters controlling e.g. the galactic winds speed and energy or black hole feedback. 
From these simulations, one can extract catalogs of galaxies with their astronomical properties such as gas mass, stellar mass, metallicity, etc. Using these catalogs, \cite{onegalcosmo, onegalastrid} showed that a machine learning model could successfully learn to predict \om{} from the features of an individual galaxy with a $\sim 10\%$ precision. %
Thus, within the context of a given suite of simulations, it seemed that the information of a single galaxy is sufficient to characterize \om, ``marginalizing'' over the other astrophysical processes modeled in CAMELS. One astrophysical parameter, $A_{SN1}$, also seemed predictable but to lesser precision. The ML approaches tested could not predict the other parameters in the same way. As pointed out by \cite{onegalcosmo}, this result indicates that the ``galaxy properties live in manifolds that change with the value of \om'' --or that at least a subset of the galaxy's properties are significantly impacted by \om. \cite{Chawak2024ApJ} extended this line of work to a slightly larger number of galaxies (up to 10), and \cite{Cosmo1gal_Chang} applied these ideas to real data using galaxy photometry observations for cosmological parameters inference but using a much larger set of galaxies ($\sim 14,000$).

However, \cite{onegalastrid} and \cite{Chawak2024ApJ} showed that the performance can degrade when transferring between different suites, i.e. training on one and testing on another. For instance, in \cite{onegalastrid} (see Figure 4), a model trained on \texttt{SIMBA} performed significantly worse on both \texttt{IllustrisTNG} and \texttt{ASTRID}; \texttt{ASTRID} seemed to provide the best results (or the least degradation) when used as a training set, but still exhibited a bias in prediction on \texttt{IllustrisTNG} (consistently over-predicting). Training on \texttt{IllustrisTNG} seemed to lead to performances that are not as biased but with larger errors and scatter. More important degradations were observed in \cite{Chawak2024ApJ} using more than one galaxy (specifically 10): in this work, while the prediction precision within a suite was significantly improved by using more than one galaxy, the performance when transferring to different suites was even more degraded than in the single galaxy case. 
These results indicate some distributional differences between suites in terms of galaxy population properties, which is expected since those suites were run using different codes that have different astrophysics implementations. For instance, while \texttt{IllustrisTNG} and \texttt{SIMBA} use a fixed blackhole seed mass (respectively $8 \times {10}^5 h^{-1} M_{\odot}$ and $1 \times {10}^4 h^{-1} M_{\odot}$), in \texttt{ASTRID} the blackhole seed mass is drawn from a power-law sampling masses in the range ($3 \times {10}^4 h^{-1} M_{\odot}$, $3 \times {10}^5 h^{-1} M_{\odot}$). The stellar mass threshold adopted in SIMBA for the BH seeding is also significantly bigger than the one adopted in IllustrisTNG  \citep{Bird_ASTRID22}. Another possibly important aspect is that \texttt{SIMBA} and \texttt{IllustrisTNG} have been calibrated using observational summary statistics and scaling relation at $z=0$ \citep{Pillepich_2018MNRAS}, while \texttt{ASTRID} simulations have been calibrated only against the observed UV galaxy luminosity function at $z=4$  \citep{Bird_ASTRID22}. 

 These discrepancies might impact inference results: \cite{casco2023} showed that the constraints obtained for \om{}, using the SPARC sample with the structural properties and dark matter content of star-forming galaxies, differed widely between suites. Some of these derived constraints were inconsistent with current constraints from other observables: \texttt{IllustrisTNG} gives \om{}$=0.27$, \texttt{ASTRID} gives \om{}$=0.44$, and  \texttt{SIMBA} \om{}$=0.14$. Their work also showed that the fiducial runs based on SIMBA and ASTRID suites predict that, at fixed stellar mass, masses of the central dark matter haloes are larger than those observed. This implies that for a given stellar mass, galaxies in \texttt{ASTRID} will reside in more massive haloes compared to \texttt{IllustrisTNG}.  Since we expect galaxies at a similar mass to reside in more massive haloes when \om{} increases \citep{lue2025iobonegal}, this is a factor, among others, that will impact a potential inference on \om{}. 
Understanding and quantifying these differences are thus crucial to improving our fundamental comprehension of how numerical implementations and modelling choices impact galaxy formation and evolution and to identifying potential systematic biases that could impact subsequent methods. In this spirit, \cite{lue2025iobonegal} recently presented a follow-up analysis of \cite{onegalcosmo} results, using Information-Ordered Bottleneck auto-encoding to investigate the effects of parameters on the reconstruction, putting in perspective scatter from the cosmic variance at a fiducial setting and the variance or effects induced by parametrization change.

Some of the concerns about the results of \cite{onegalcosmo} and follow-up works can be summarized with the following key points: first, galaxy formation is far from being completely resolved, and the assumptions made in the hydro-simulations are known to be approximated at most; additionally, incomplete or incorrect sub-grid physics, as well as the limitations of simulations' volumes, could impact our conclusions regarding the real universe, as seen in e.g. \cite{casco2023}. Second, the impact of the priors used when building the simulations (e.g. choice of astrophysical parameters, their ranges, etc) could be the main driver of the observed results --this aspect overlapping with the previous point, since one could see the choices of hydro-simulations as ``priors'' as well. These aspects would prevent using galaxy properties as a reasonable cosmological probe in practice.

In line with these arguments, and following the initial deduction from \cite{onegalcosmo} --and subsequent work from \cite{lue2025iobonegal}-- that \om{} induces changes in the galaxy properties ``manifold'', we propose to investigate the statistical properties of the simulations in the CAMELS suite, across cosmological parameters and (some) hydrodynamic choices, to clarify the underlying mechanisms behind the results.
We also aim to provide preliminary physical hypotheses explaining these properties from a cosmological perspective. 
We focus on the following key questions: 

\begin{itemize}
\item What are the respective effects of the cosmological and astrophysical parameters on the distributions of galaxy properties? That is, can we quantify the ``signal'' a parameter introduces in the resulting population, both in terms of strength and in terms of nature?
\item Do we identify distributional degeneracies, i.e. parameters confounders for \om{}? 
\item What impact do hydro-simulation choices have on galaxy properties and on the effect of the parameters in this context? For instance, is the effect of \om{} consistent across suites?
\end{itemize}

We introduce the simulations used in Section \ref{sec:data}.
In Section \ref{sec:paramimpact}, we investigate the global ``strength'' of the effect of each parameter on the distribution of galaxy properties, using simple binary classification methods to estimate how separable two simulations are. 
This analysis shows that some parameters do not impact the distribution of galaxy properties at all and thus can easily be marginalized over. However, $A_{SN1}$ has an effect of similar strength as \om{}. We further explore the effects of parameters, now combined together, in Section \ref{sec:degeneracies}: we look at the distribution of likelihood estimates in parameter space, i.e. measuring how much galaxies from a simulation are ``probable'' in different simulations. Those show that $A_{SN1}$ introduces distributional degeneracies for \om{} in a subset of the feature space, but these degeneracies can be disentangled using additional features. In Section \ref{sec:astrid}, we use Optimal Transport to measure the displacement vector induced by a change in parametrization: this provides a linear estimate of the effect of a parameter on each feature. We compare \texttt{IllustrisTNG} and \texttt{ASTRID}, investigating the impact of the hydro-simulation used on the displacements induced by parameters and on the distributions of galaxy properties at fixed parametrization. The overall trend of \om{}'s displacement seems coherent with the physical model of cosmology and relatively consistent across the two suites when increasing \om{}; decreasing \om{} exhibits more discrepancies between the different hydro-codes. However, those differences are hard to disentangle from the distributional discrepancies that are also observed (as expected) between the two suites for similar parametrization (for instance in metallicities and radius features). 
We discuss our results in Section \ref{sec:discu}. 
We also provide in the Appendix \ref{sec:boundaries} an analysis of the impact of the Latin Hypercube boundaries range (that is, our priors on the parameters ranges) on the performance and predictability of \om. Appendix \ref{sec:OTviz} contains additional illustrations and considerations regarding the Optimal Transport approach.

\section{Simulations}
\label{sec:data}

We first focus our analysis on \texttt{CAMELS-IllustrisTNG} \citep{camels, camels_2023}, a suite of hydrodynamic simulations using the moving-mesh AREPO code \citep{springel2010arepo,weinberger2020arepo} and based on the \texttt{IllustrisTNG} framework \citep{2017MNRAS.465.3291W,2018MNRAS.475..676S,2018MNRAS.473.4077P,2018MNRAS.475..624N,2018MNRAS.480.5113M}. We will refer in the remainder of the paper to the simulations from this suite in \texttt{CAMELS} simply as \texttt{IllustrisTNG} for short.

In Section \ref{sec:astrid}, we also compare \texttt{CAMELS-IllustrisTNG} with \texttt{CAMELS-Astrid} \citep{CAMELSDR2} suite, which uses a refined version of MP-GADGET3 \citep{feng2018mp} and the \texttt{ASTRID} model \citep{Bird_ASTRID22, ni2022astrid}. This suite shares a subset of parameters with similar associated ranges as \texttt{IllustrisTNG}, but the implementation of the astrophysical parameters are in themselves different.

Specifically, we use the ``Latin Hypercube'' (LH) suite, which varies, for \texttt{IllustrisTNG}, 2 cosmological parameters, \om{} and $\sigma_8$, and 4 astrophysical parameters, $A_{SN1}$ (galactic winds energy), $A_{AGN1}$ (black hole feedback kinetic mode energy), $A_{SN2}$ (galactic winds speed), $A_{AGN2}$ (black hole feedback kinetic mode burstiness/energy speed). The astrophysical parameters are uniformly sampled in their log10 within the boundary of their prior. The ranges for the Latin hypercube sampling of the parameters are, for \texttt{IllustrisTNG}:
\begin{equation}
    \begin{split}
        \Omega_{\rm m} & \in [0.1, 0.5] \\
        \sigma_8 & \in [0.6, 1.0] \\
        \log_{10}(A_{SN1}), \log_{10}(A_{AGN1}) & \in [-0.6, 0.6]\\
        \log_{10}(A_{SN2}), \log_{10}(A_{AGN2}) & \in [-0.3, 0.3]
    \end{split}
\end{equation}
In the rest of the paper and the figures, we report the astrophysical parameters' values,   \textit{after} log (i.e. within the ranges described above); for notation lightness, we do not explicitly use $log_{10}(A)$ but simply $A$. \texttt{ASTRID} simulations follow the same ranges, except for $\log_{10}(A_{AGN2})$ which range from $[-0.6, 0.6]$; see \cite{2021MNRAS.507.1999H} for additional informations on \texttt{ASTRID} specificities.

The fiducial simulation is considered at the centre of the parameter space, that is $\Omega_{\rm m}=0.3, \sigma_8=0.8, A_{SN1}=A_{AGN1}=A_{SN2}=A_{AGN2}=0$. 

We also use the 1P simulations, available for both \texttt{IllustrisTNG} and \texttt{ASTRID}, which vary only one parameter at a time --the others being fixed at their fiducial values. This suite provides, for each parameter, 5 simulations where the parameter takes the value of the 0-th, 25-th, 50-th (fiducial), 75-th and 100-th percentile of its range.

For each simulation, a catalog of galaxies is extracted, where the halos and subhalos are identified using \texttt{SUBFIND} \citep{springel2001popcluster,dolag2009substruct}. Each galaxy has the following properties: 
\begin{enumerate}
    \item \textbf{$M_g$}: gas mass content of the galaxy, including circumgalactic medium.
    \item \textbf{$M_*$}: stellar mass.
    \item \textbf{$M_{BH}$}: black hole mass.
    \item \textbf{$M_t$}: total mass, i.e. sum of dark matter, gas, stars, and black-holes masses in the subhalo.
    \item \textbf{$V_{max}$}: maximum circular velocity of
the subhalo hosting the galaxy: $V_{max} = max(\sqrt{GM(<R)/R})$
    \item \textbf{$\sigma_v$}: velocity dispersion of all particles contained in the galaxy’s subhalo.
    \item \textbf{$Z_g$}: mass-weighted gas metallicity of the galaxy.
    \item \textbf{$Z_*$}: mass-weighted stellar metallicity of the galaxy.
    \item \textbf{$SFR$}: galaxy's star formation rate.
    \item \textbf{$J_{spin}$}: modulus of the galaxy’s subhalo spin vector.
    \item \textbf{$V_{pecu}$}: modulus of the galaxy’s subhalo peculiar velocity.
    \item \textbf{$R_*$}: radius containing half of the galaxy stellar mass.
    \item \textbf{$R_t$}: radius containing half of the total mass of the galaxy’s subhalo.
    \item \textbf{$R_{max}$}:  radius at which $\sqrt{GM(<R_{max})/R_{max}}=V_{max}$
    \item \textbf{$U$}: galaxy magnitude in the U band.
    \item \textbf{$K$}: galaxy magnitude in the K band.
    \item \textbf{$g$}: galaxy magnitude in the g band.
\end{enumerate}

We consider in this work galaxies with $M_* > 5 \times 10^8 h^{-1} M_{\odot}$, similar to \cite{onegalastrid}.

We do not include galaxy magnitudes in our analyses since these features are available only for the \texttt{CAMELS-IllustrisTNG} simulations and not for \texttt{CAMELS-Astrid}. In the following analyses, we use the values after $\log_{10}$ for the mass features (i.e., $M_*, M_{BH}, M_g, M_t$).

We denote by $\Sfeat{i}{M}$ the vector of properties of galaxy $i$ from simulation $S_M$ with parameters $\params^M$, and by $\Gfeat{M}$ the set of galaxy feature vectors from simulation $S_M$, i.e. $\Gfeat{M} = \{\Sfeat{1}{M}, \dots, \Sfeat{g_M}{M} \}$, where $g_M$ is the number of galaxies in simulation $S_M$. In the rest of this paper, we refer to $\Sfeat{i}{M}$ as `galaxy properties' in the context of their physical impact, and as `galaxy features' (or simply `features') when looked at from a machine learning perspective.

\section{Global individual impact of parameters}
\label{sec:paramimpact}

From a machine learning perspective, the results from \cite{onegalcosmo} imply not only that the properties of a galaxy can be informative of \om{} but also that we can marginalize over the other parameters of the suite when predicting \om. This marginalization can occur under two scenarios: if considering a system with two parameters $A$ and $B$ where we observe $A$ being predictable, either (i) parameter $B$ has a minimal impact on the features, or an impact much smaller than $A$'s, making marginalization straightforward, (ii) parameter $B$ affects the features but does not introduce strong degeneracies (or enough variance) that would obfuscate $A$'s signal.

To understand in which scenario we are, we propose to use the 1P-suite of \texttt{CAMEL-IllustrisTNG} to evaluate the global effect each individual parameter has on galaxy properties. 
The one-dimensional shift in parameter space provided by the 1P simulations enables us to check, along each direction in parameter space, how different and distinguishable the distributions of the galaxy properties are between two simulations along that direction.

To quantify the difference or overlap between the distributions, we propose a simple and intuitive test using the accuracy of a classifier trained to distinguish two simulations from a single galaxy -- a simplified version of the original parameter prediction setup. Consider $\theta_{(k)} \in \params$ the parameter's coordinate being varied (e.g. \om), and two values $\theta_{(k)}^L$ and $\theta_{(k)}^M$ of this parameter, leading to simulation $S_L$ and $S_M$, respectively. We assign a binary label to each simulation, and we train a classifier (RandomForestClassifier from \texttt{sklearn} \citep{scikit-learn} with default settings) to predict the label of the correct source simulation given the features of a galaxy, $\Sfeat{i}{J}$, where $J={L,M}$.

We take 70\% of the data (without re-balancing) to train and 30\% as a test set to compute the balanced accuracy (defined as the mean between the True Positive Rate and the True Negative Rate, to account for the potential differences in the number of galaxies between two simulations). We repeat this operation 50 times and report the mean and variance of the resulting test accuracies. We note that we obtain similar results when using rebalanced datasets, correcting for the difference in the number of galaxies in each simulation.

We conduct this distinguishability test for the simulations along the range of a single parameter $k$ against the fiducial simulation. When comparing the fiducial simulation against itself (which is, namely, a baseline-check for indistinguishability), we randomly split the dataset in half to assign the binary labels. If the distributions of galaxy features under $\theta_{(k)}^L$ and $\theta_{(k)}^M$ are very similar, the classifier's accuracy should be near random. The more the distributions depart from each other (i.e. the less there is an overlap), the higher we expect the accuracy to be, as a classifier can easily distinguish them. 

\begin{figure*}
    \centering
    \includegraphics[width=\linewidth]{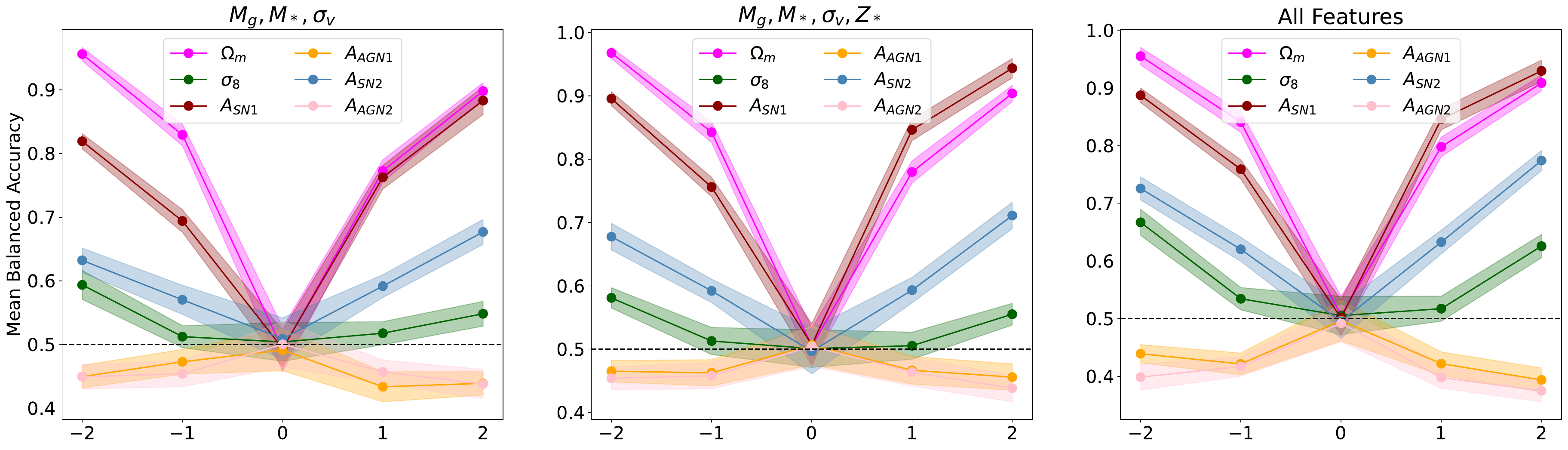}
    \caption{Mean (and standard deviation) of balanced accuracy (i.e. mean between true positive rate and true negative rate) across 50 runs of classifiers trained to distinguish between galaxies from the fiducial simulation (0 label on the horizontal axis) and a different simulation with a single parameter value changed (indicated by the color), using \texttt{CAMEL-IllustrisTNG}. The magnitude of the parameter difference is given by the horizontal axis (-2 being the minimal value of the parameter's range, -1 the 25th percentile, 0 the fiducial, 1 the 75th percentile, 2 the maximum value taken by the parameter). The left panel shows results using only three features ($M_g, M_*, \sigma_v$), the middle panel includes the additional feature $Z_*$, and the right panel uses all features. }
    \label{fig:accuracyClassif_1P}
\end{figure*}

Figure \ref{fig:accuracyClassif_1P} shows the mean accuracy (and its standard deviation) as a function of $\theta_{(k)}^L$, i.e. the parameter's value for the simulation $S_L$ being classified against the fiducial (where, on the horizontal axis, $-2$ indicates the minimal value in the range, $-1$ is the 25th percentile, $0$ the fiducial (50th percentile), $1$ the 75th percentile, and $2$ the maximum value in the range). The color indicates the parameter being varied. The three panels of Figure \ref{fig:accuracyClassif_1P} show accuracies calculated using an increasing number of features: from three features, $M_g, M_*, \sigma_v$ -- selected following  \cite{onegalcosmo} and \cite{Chawak2024ApJ}, who found that \om{} had a strong impact on $M_*$ and $V_{max}$, which is highly correlated with $\sigma_v$-- to four features --  
including also $Z_*$ (one of the top four features impacted by \om{} in the work of \cite{onegalcosmo, Chawak2024ApJ}), and, finally using all features.

The plot shows that simulations with different \om{} values (pink) are easily distinguishable from each other (high accuracy in classification). The same is true for $A_{SN1}$ (brown). These parameters impact (some of) the features' distributions significantly: the accuracies indicate an overlap of distributions of around 10\% (as a 90\% accuracy indicates that 10\% of the galaxies cannot be correctly distinguished). The effect is asymmetric: higher \om{} influences the distribution less than lower \om{}, as demonstrated by its lower accuracy; we observe the opposite effect for $A_{SN1}$. This is consistent with the results observed in \cite{lue2025iobonegal}. $A_{SN2}$ and $\sigma_8$ both have some impact, but more so in the remaining features (third panel). $A_{AGN1}$ and $A_{AGN2}$ show very little influence on the distribution, as it is impossible to distinguish between them. We note that there are other ways to estimate the difference between two distributions. For instance, \cite{hagen2021accelerated} proposed an extension of the Kolmogorov-Smirnof distance to multi-variate data. We found similar trends using this approach.

From this test, we conclude the following:
first, $A_{AGN1}$ and $A_{AGN2}$ do not affect the distribution of the simulations' features, thus making marginalizing over these parameters trivial; at the same time, this also explains why those parameters were found to be not predictable by \cite{onegalcosmo}, a consequence of the lack of signal in the feature space. It is interesting to note that \cite{2024arXiv241204559L} also noted a lack of effect of the AGN parameters in a different context. Second, $\sigma_8$ and $A_{SN2}$ have a limited impact on the reduced sets of features, but seem to affect some of the remaining features (right-most panel), albeit still to a much lower magnitude than \om. The effect of $\sigma_8$ seems non-linear, as indicated by the trend in accuracy for its extreme values compared to the 25th and 75th percentile (-1 and 1 on the x-axis). Finally, $A_{SN1}$ has a strong impact on the distribution of galaxy features, of a comparable magnitude as \om{}, even in our reduced sets of features.

\section{Estimated Likelihood and Potential Distributional Degeneracies}
\label{sec:degeneracies}

We now investigate the 
distribution of the properties when varying several parameters simultaneously. The previous analysis showed that $A_{SN1}$ has a strong effect on the distribution of galaxy properties. Additional experiments illustrate that, under certain conditions, $A_{SN1}$ can deteriorate the predictability of \om{} (see Appendix \ref{sec:boundaries}). This could happen if $A_{SN1}$ increases the variance of the features, or if changing $A_{SN1}$ shifts the distribution in a way that can be counteracted by adjusting \om{}. We refer to the latter effect as `distributional degeneracy'.
 
 We implement the following procedure: we estimate the distribution $P(\Sfeat{i}{J} | \params^J)$, where $\Sfeat{i}{J}$ denotes the features of galaxy $i$ in simulation $S_J$ (with parameters $\params^J$), with a Kernel Density Estimator (KDE). This estimator can then be used to evaluate the PDF of galaxies coming from a different simulation, where  $P(\Sfeat{i}{K} | \params^J)$ gives the probability density estimate of galaxy $\Sfeat{i}{K}$ (which is obtained in a simulation $S_K$ with parameters $\params^K$) within the sample of galaxies of simulation $S_J$ with parameters $\params^J$, which we call ``estimated likelihood''. 
 
 We average the estimated likelihood over all the galaxies in simulation $S_K$, obtaining the ``mean estimated likelihood'' of the simulation $K$ for $\params^J$:
\begin{equation}
    \avlike{K}{J} = \frac{1}{|S_K|}\sum_{\Sfeat{i}{K} \in S_K} P(\Sfeat{i}{K} | \params^J)\, .
\end{equation} 
To ensure that all the $P(\Sfeat{}{K}|\params^J)$ have the same normalization across different $J$, we fit the KDEs using a random subsampling of 100 galaxies from the simulation being fitted ($S_J$). Additionally, the KDE fits are computed on a standardized feature space.

We can thus pick a ``reference'' simulation, and compute its mean estimated likelihood for all the other simulations, obtaining a score for each parameter configuration covered by the Latin Hypercube sampling. A large mean estimated likelihood in some region of parameter space indicates that a large fraction of the galaxies in the reference simulation lie in a high density (i.e., probable) region of the simulation $S_J$ that is being compared; when this is the case, a machine learning algorithm will struggle to provide a good estimate for the galaxies in the reference simulation, as they are similarly distributed to the ones in $S_J$, acting as confounders. 
Conversely, a low mean estimated likelihood indicates a difference between the two distributions at different parameter values, which in turn means that an ML algorithm might predict the parameter values more easily and precisely.

\begin{figure*}
    \centering
\includegraphics[width=0.8\linewidth]{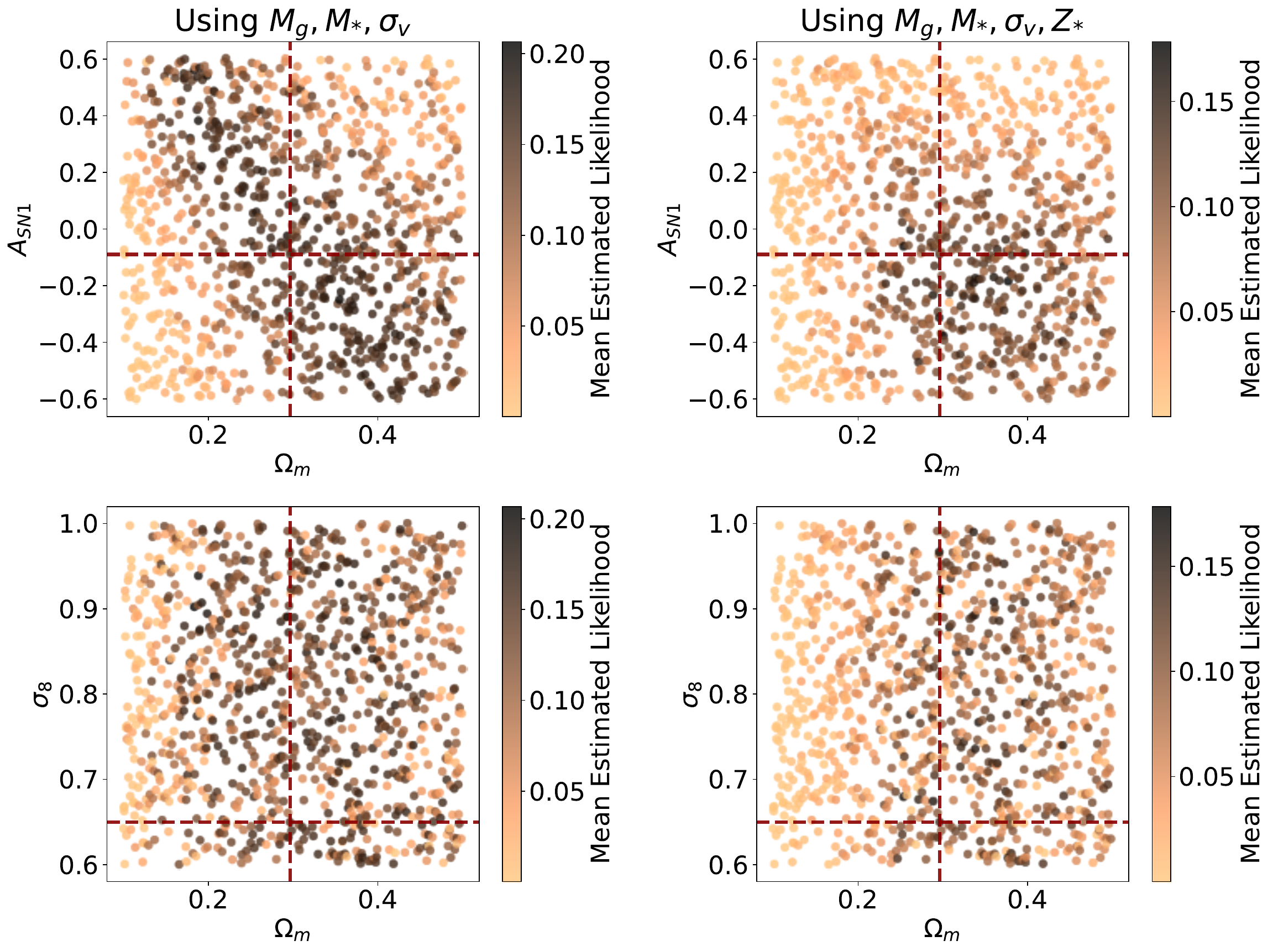}
    \caption{Mean estimated likelihood of the galaxies of a reference simulation $S_K$ (at the intersection of the dashed red lines) using the feature distribution of another simulation, $S_J$, using $M_g, M_*, \sigma_v$ (first column) and adding $Z_*$ (second column). This is computed with simulations of \texttt{CAMEL-IllustrisTNG}. }
    \label{fig:likelihood1}
\end{figure*}

\begin{figure*}
    \centering
    \includegraphics[width=0.8\linewidth]{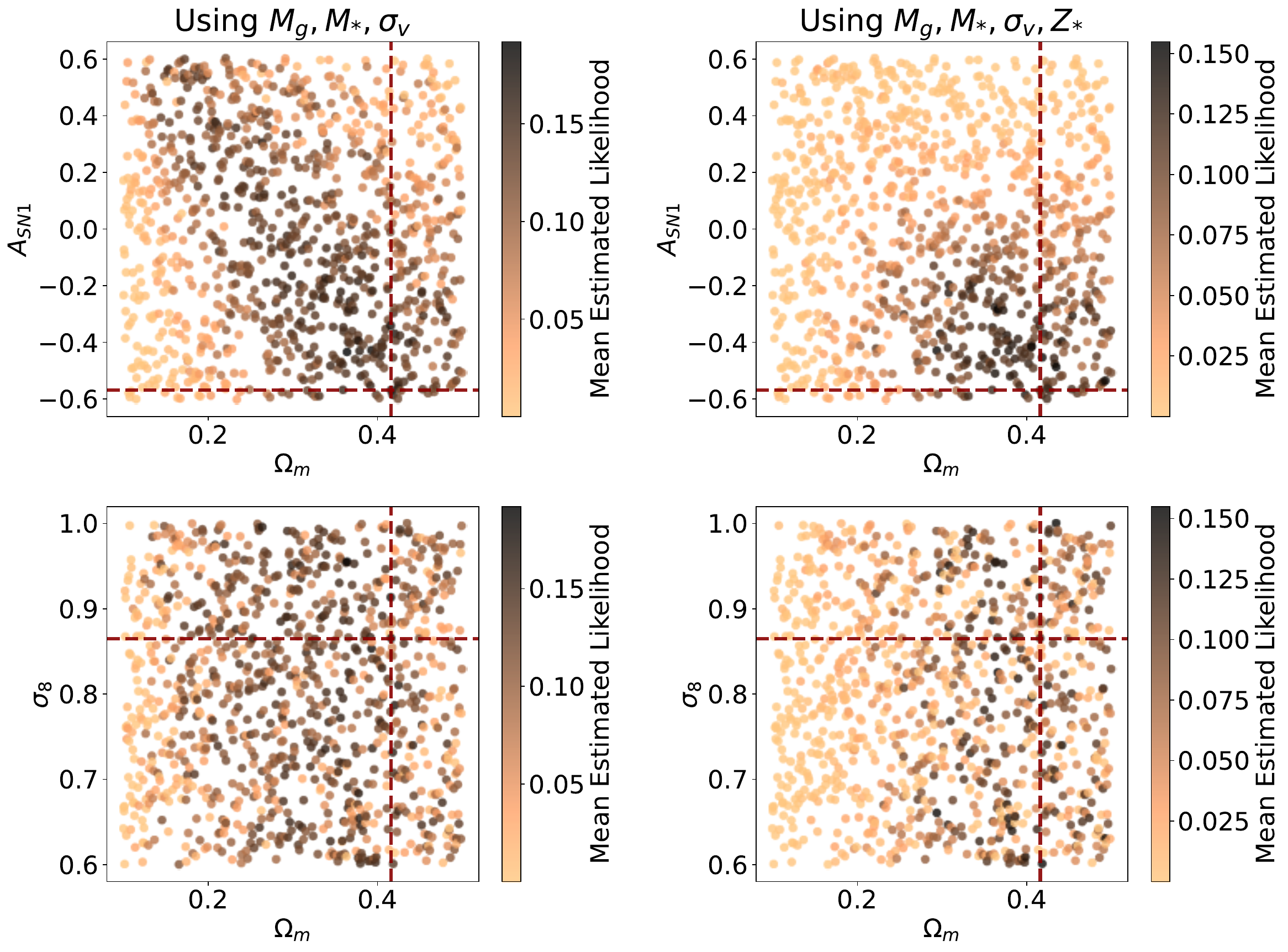}
    \caption{As in Figure~\ref{fig:likelihood1}, but for a different reference simulation.}
    \label{fig:likelihood3}
\end{figure*}

We plot in Figure \ref{fig:likelihood1} the mean estimated likelihood of a reference simulation $S_K$ (whose parameters $\theta^K$ are indicated by the dark-red dashed lines) for all other simulations in \texttt{Illustris}-LH (thus varying all parameters), in different projections of the parameter space. Figure \ref{fig:likelihood3} is constructed in the same way but using a different reference simulation. The first columns of each figure display the value of the mean estimated likelihood computed using only the three features $M_g, M_*$ and $\sigma_v$. The second column is computed including the additional feature $Z_*$\footnote{We are restricted to a relatively small number of features as the KDE computation can become unreliable in the full, high-dimensional feature space due to the usual curse of dimensionality.}.  

Using only 3 features, we note the existence of a region in the \om{} and $A_{SN1}$ space, running diagonally in an anti-correlated manner, where galaxies from $S_K$ (the reference simulation) lie in a high-density region of the feature spaces for simulations $S_J$ (region with large values of $\avlike{K}{J}$). In other words, the bulk of the galaxy features from the reference simulation is similar to --or at least is significantly contained within-- simulations with a different \om{} value when $A_{SN1}$ is changed in an anti-correlated way with \om{}. This region thus constitutes a distributional degeneracy. Interestingly, this can also be identified in \cite{lue2025iobonegal}, Fig. 3. This effect is seemingly not impacted by the other parameters' values, as it would be otherwise obfuscated in the projection. We also do not see comparable trends in other projections in the parameters space. We would not expect to see any distributional degeneracies as a function of $A_{AGN1}$ or $A_{AGN2}$ since those parameters were identified in Section \ref{sec:paramimpact} as having barely any impact on the distribution of galaxy properties. However, the distributional degeneracy between $A_{SN1}$ and \om{} is surprising: this requires the two parameters to have very similar (if opposite) effects in terms of direction and magnitudes (see Section \ref{sec:astrid} for more visualizations on that regard).

\begin{figure*}
    \centering
    \includegraphics[width=.8\linewidth]{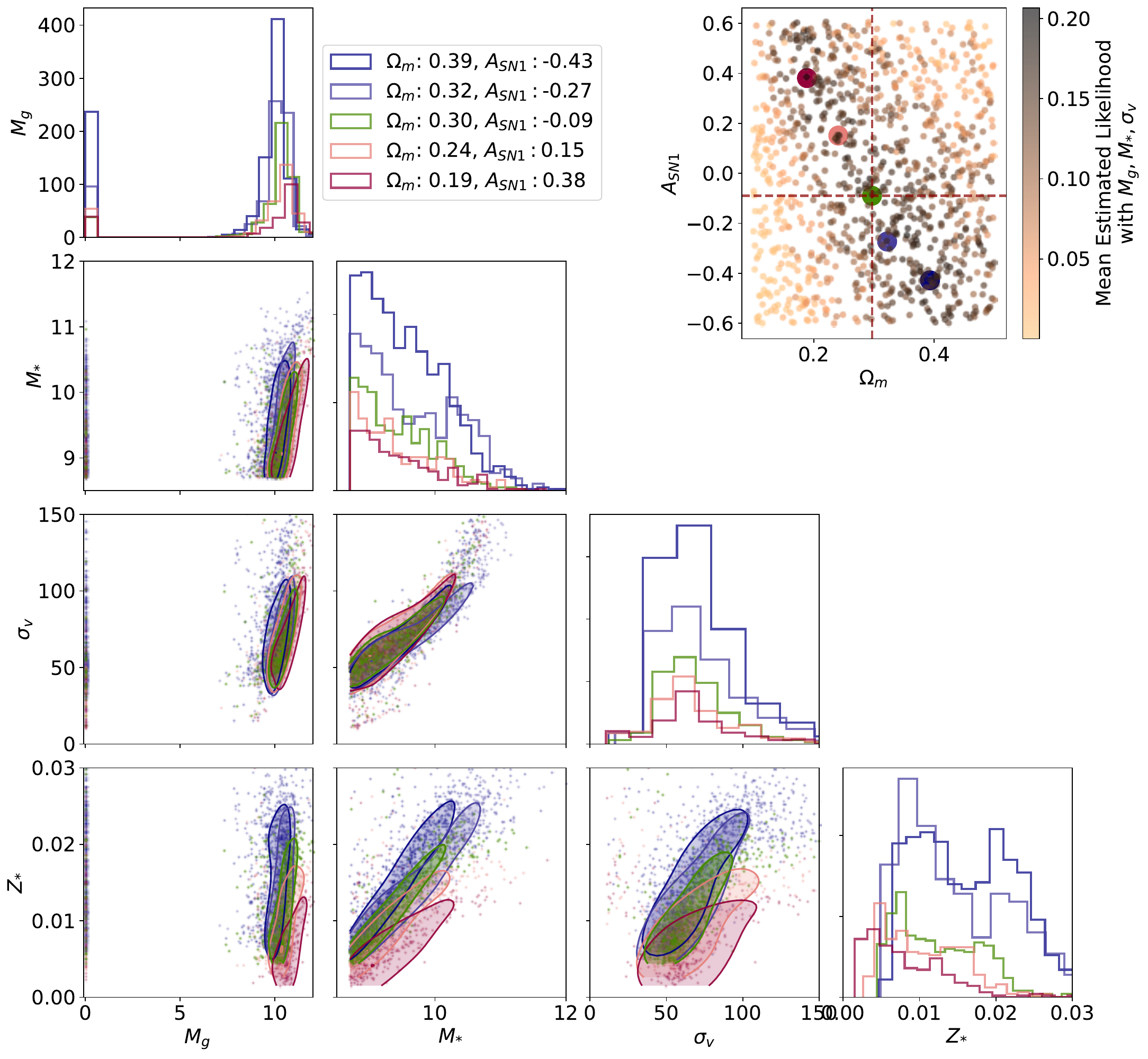}  
    \caption{Visualization of the distribution of the galaxies' features for 5 simulations from \texttt{CAMEL-IllustrisTNG} --location in parameter space shown in the top right panel, with all other simulations colored by their mean PDF as shown in Fig \ref{fig:likelihood1}. The target simulation is shown in green. The contours show the 30th percentile of the estimated densities --computed in the four features space, i.e., tracing around 70\% of the data points.}
    \label{fig:blob_targ0}
\end{figure*}

Adding $Z_*$ removes some of the distributional degeneracy for high $A_{SN1}$: the mean estimated likelihood drops significantly for simulations with lower \om{} and high $A_{SN1}$, leading to a tighter distribution of the estimated likelihood around the actual \om{} value. This would, in turn, translate into a more precise estimation of \om{}. We can visualize these distributional degeneracies (and how the addition of $Z_*$ breaks them) by picking a few simulations along the distributional degeneracy direction. We plot in Figure \ref{fig:blob_targ0} the distribution of the reference simulation (in green) and of 4 other simulations, in a corner-plot style, showing 2-dimensional projections of the 4-D feature space. The contours are computed to trace the 30th percentile of the density estimate (computed in 2D), i.e. they encompass 70\% of the galaxies in the simulation.
Note that the number of galaxies in a simulation varies (also as a function of $\Omega_{\rm m}$), which is reflected in Figure \ref{fig:blob_targ0}: the dark purple distribution ($\Omega_{\rm m}\sim 0.39$) has many more galaxies than the dark red one ($\Omega_{\rm m}\sim 0.19$). We observe that the contours for the reference simulation (green) in $M_g,  M_*$, and $\sigma_v$ largely overlap with those from the other simulations picked along the degeneracy direction. This means that typical galaxies from the reference simulation are typical for the other simulations, too --even though e.g. the distribution for $\Omega_{\rm m}=0.39, A_{SN1} = -0.43$ (dark purple) extends to regions where the green distribution does not reach. However, when considering $Z_*$ (bottom row), the distribution for the dark red simulation is significantly 
shifted away from the reference simulation (green). This makes galaxies extracted from the reference simulation less likely for a simulation with parameters $\Omega_{\rm m}=0.2, A_{SN1} = 0.41$, 
and thus breaks the distributional degeneracy, as observed in Figure \ref{fig:likelihood1} in the top-right panel. This serves as an intuitive visualization of how adding features --and given a model robust to potentially noisy or useless features and able to scale to higher dimensional data-- should only improve the likelihood estimate (as we gain information) and, in turn, the precision of the prediction. These experiments also seem to indicate that there might be, indeed, a peak in the likelihood. 

\section{Comparing parameters' effects in \texttt{IllustrisTNG} and \texttt{ASTRID}}
\label{sec:astrid}

The analyses from Sections \ref{sec:paramimpact} and \ref{sec:degeneracies} highlight the following:

\begin{itemize}
    \item \om{} has a strong impact on the distribution of galaxies in feature space, leading to little ($\sim 15\%$) distributional overlap in at least a subset of the features when varying \om{} by $\pm 0.1$.%
    \item Several astrophysical parameters have no effect on the distribution of galaxy properties. 
    \item Other parameters have an effect, $A_{SN1}$ specifically having a signal of similar strength as \om. While this creates some distributional degeneracies with \om, they are restricted to a subset of the feature space. 

\end{itemize}

These findings explain, from the statistical properties of the simulations perspective, that \om{} can be predicted to $\sim 10\%$ precision on average from a single datapoint of a simulation. 

However, this does not tell us, from a physics perspective, {\em why }\om{} has such a strong impact --or if it shouldn't; and {\em why} some of the astrophysical parameters do not have a comparable effect --or if they should.

By looking more closely at \textit{how} \om{} impacts galaxy properties, we can potentially connect it to physically grounded hypotheses (albeit a posteriori; validating those being beyond the scope of this paper). Since the role and impact of the hydro-simulations used have been debated in the wake of the original results (and are a crucial aspect for those simulations), we propose to compare \texttt{IllustrisTNG} and \texttt{ASTRID}, which shares the same set of parameters as \texttt{IllustrisTNG} but with different effective implementations, as well as a different hydrodynamic code.

Our goal is to refine the estimation of the parameters effective impacts on the distributions of galaxy properties; in other words, we want to quantify more specifically how the distribution of galaxy properties changes between two different simulations to evaluate, for instance, if increasing one parameter creates on average galaxies with higher stellar mass, or changes the spread of the galaxies' velocities. 

To this end, we use Optimal Transport with the \texttt{POT} library \citep{flamary2021pot}. We consider two simulations at different parameter values, $S_N$, and $S_M$, each with its set of galaxies,  $S_N =\{\Sfeat{1}{N},\dots,\Sfeat{g_N}{N}\}, S_M = \{\Sfeat{1}{M},\dots,\Sfeat{g_M}{M}\}$, where $\Sfeat{i}{N},\Sfeat{i}{M} \in \mathbb{R}^d$ (note that the two sets can have different numbers of galaxies $g_N$ and $g_M$). We compute a distance matrix ${\bf D} \in \mathbb{R}^{g_N \times g_M}$ (here using Euclidean distance) between each element of $S_N$ and each element of $S_M$. Using this distance matrix (as a ``cost function''), the method solves the Earth Mover Distance problem using the \cite{bonneel2011displacement} algorithm to find the optimal transportation matrix $\boldsymbol \gamma \in \mathbb{R}_+^{g_N \times g_M}$:

\begin{equation}
    \begin{split}
        \boldsymbol \gamma &= \operatorname{arg min}_{\boldsymbol \gamma} \langle{\boldsymbol \gamma}, {\bf D}\rangle_F 
    \\
\text{such that} &  \begin{cases}
{\boldsymbol \gamma} \vb{1}^{g_M}  = \vb{a}, \\
         {\boldsymbol \gamma}^T\vb{1}^{g_N}  = \vb{b}, \\
         {\boldsymbol \gamma} \geq 0 
\end{cases}
\end{split}
\end{equation}

where $\vb{a}$ and $\vb{b}$ are the sample weights (chosen as uniform weights in our case) for each set, i.e. $\vb{a} \in \mathbb{R}^{g_N}, \vb{b} \in \mathbb{R}^{g_M}$; $\vb{1}^{g}$ is column vector of ones of size $g$, and $\langle \cdot,\cdot \rangle_F$ is the Frobenius inner product.   

Intuitively, $\boldsymbol\gamma$ tells us how to move (or re-express) the points in $S_N$ as a function of the points of $S_M$. $\boldsymbol \gamma_{i,j}$ expresses how much $\vb{x}_j^M$ ``weight'' in the transport of $\vb{x}_i^N$ onto $S_M$. Minimizing the Frobenius inner product between $\boldsymbol\gamma$ and the distance matrix encourages ``matching'' together points that are not too far from each other, and dropping $\boldsymbol\gamma_{i,j}$ to 0 if ${\bf D}_{i,j}$ is large. The conditions on $\vb{a}$ and $\vb{b}$, on the other hand, enforce that each point has at least one matching counterpart, avoiding trivial solutions like $\boldsymbol\gamma$ being null or some points being discarded. 
The entire matrix $\boldsymbol\gamma$ sums to one (when $\vb{a}$ and $\vb{b}$ are defined as vectors summing to one), and we propose to use it to compute the global average displacement vector as:
\begin{equation}
    \overrightarrow{ST} = \sum\limits_{j=1}^{g_N} \sum\limits_{i=1}^{g_M} \boldsymbol\gamma_{j,i}(\Sfeat{j}{M}- \Sfeat{i}{N})
\end{equation}
While this is equivalent to computing the difference between each set's means (by construction of $\vb{a}$ and $\vb{b}$), the optimal transportation matrix $\boldsymbol\gamma$ does provide additional information by ``linking'' galaxies between the simulations. This allows us to compute, for instance, the standard deviation over the individual average displacement vectors. This is relevant, for the average displacement vector can be uninformative in some cases: for instance, when the mean of both distributions remains the same but one variance increases (e.g., if one set is sampled from $ N(0,1)$ and another from $ N(0,2)$). We provide in Appendix \ref{sec:OTviz} a visualization of the optimal transport linkage and the resulting vectors on two simulations.

\begin{figure*}
    \centering
    \includegraphics[width=.9\linewidth]{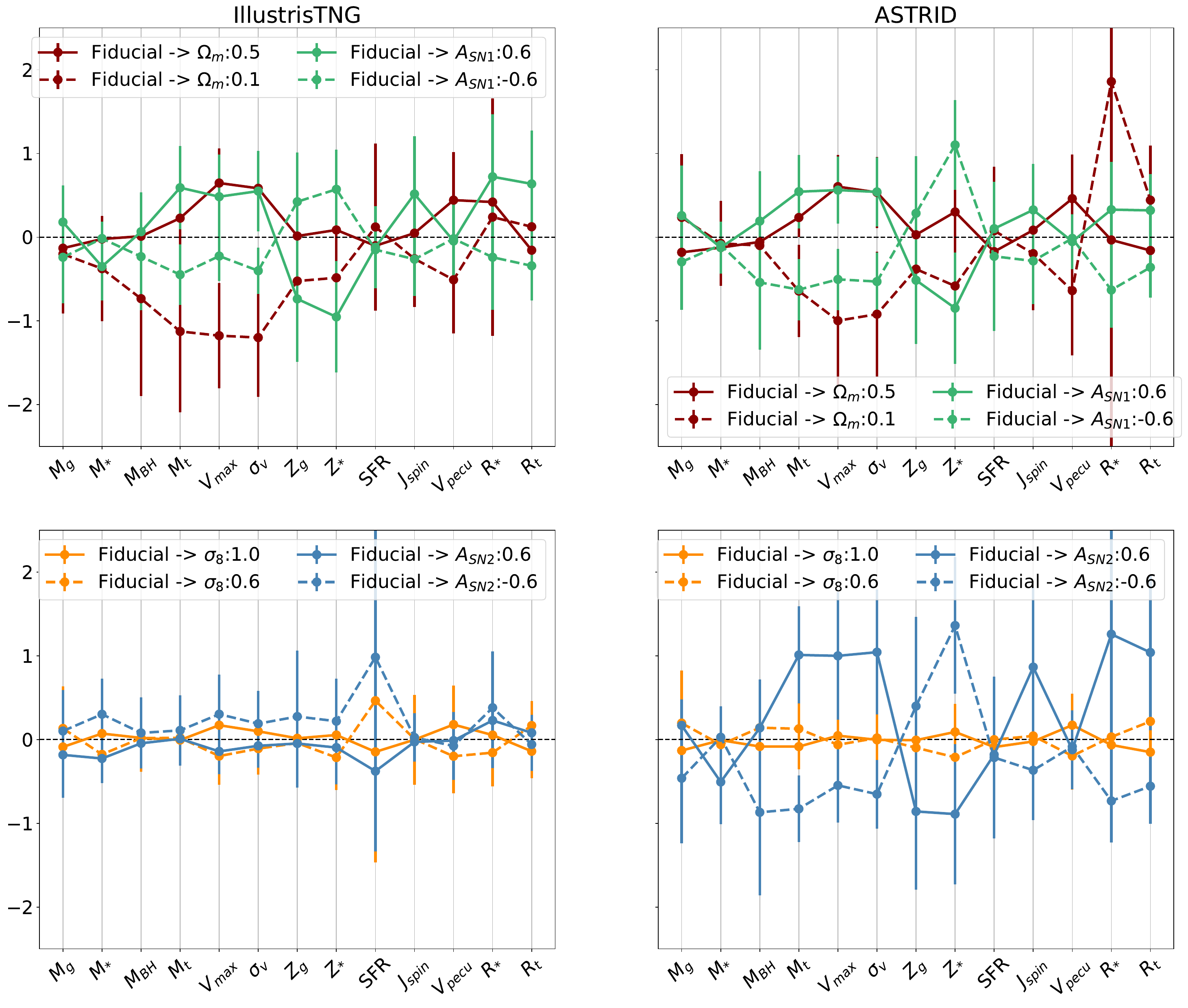}
    \caption{Average Displacement Vector, computed using Optimal Transport and the Earth Mover Distance algorithm, between the fiducial distribution and different target distributions. Each simulation dataset was standardized with respect to the fiducial simulation before computing the Earth Mover Distance; the y-axis shows the value of the ADV in each dimension in the normalized space. The color indicates the parameter changed (\om{} in dark red $A_{SN1}$ in sea-green, $\sigma_8$ in orange, $A_{SN2}$ in blue); the line-style indicates the value of the parameter, where solid lines equates to the maximum value of the parameter and dashed-lines to the minimum value. The left column shows the displacement for \texttt{IllustrisTNG}; the right column for \texttt{ASTRID}. Intuitively, this vector represents how, on average, galaxies from the fiducial simulation are displaced to match the ones from the target simulation.}
    \label{fig:OT}
\end{figure*}

We plot in Figure \ref{fig:OT} the mean and standard deviation of Optimal Transport displacement vectors computed using \texttt{IllustrisTNG}-1P in the left panels and with \texttt{ASTRID}-1P in the right panels. All displacement vectors are computed using the fiducial simulation (of the respective suite) as the first set $S_N$; the second set $S_M$ corresponds to a simulation where only a single parameter is varied. The parameter varied is indicated by the color of the line and is set either to its minimum value (dashed lines) or to its maximum value (solid), e.g. the dashed dark-red line indicates the average displacement between the fiducial simulation and the simulation with \om$=0.1$.

The displacement vector is computed using all features, i.e. the distance matrix was computed in the 13-dimensional feature space. We standardized the simulations with respect to the fiducial simulation of the respective suite, as the features have different magnitude ranges. Each line thus depicts on the vertical axis the magnitude of the average displacement for the parameter indicated on the horizontal axis needed to match the galaxies of two simulations, in a standardized space: a positive  (resp. negative) value for a feature $x$ indicates that the mean of the distribution for $x$ is larger (resp. smaller) in the simulation $S_M$. A large error bar indicates  that the direction of the individual displacement vector has a large variance, which means a difference in the variance of the two distributions. We do not show $A_{AGN2}$ as the range used is different in \texttt{ASTRID} suite (thus, extreme values are not directly comparable between the two suites). $A_{AGN1}$ showed no effect in both suites and is thus not depicted as well.

Figure \ref{fig:OT} confirms the trend observed in the previous Sections on the asymmetry of the effect of \om{} between low and high values:  the dashed dark-red line in the top panel (\om$=0.1$) indicates stronger displacement than its dark-red solid counterpart (\om{} $=0.5$). It also confirms the similarity of the effect of high \om{} and high $A_{SN1}$ for features such as $\sigma_v$ and $V_{max}$, leading to the distributional degeneracies in that subspace. There is also asymmetry in the effect of $A_{SN1}$, but it is less pronounced than for \om. Those also help explain the results obtained by \cite{lue2025iobonegal}: they observed that the reconstruction quality of galaxy properties (when using an Information Ordered Bottleneck Auto-Encoder trained on only fiducial simulations) were impacted by changes in \om, but also in $A_{SN1}$, and that errors changed in an asymmetric way (see Figure 2 in their work). Galaxies with high $A_{SN1}$ and low \om{} had higher reconstruction errors than resp. low $A_{SN1}$ and high \om{} (which themselves had higher errors compared to the fiducial). Our OT experiment with Figure \ref{fig:OT} shows that those parameters indeed change the distribution of galaxy properties significantly, thus making most galaxies of those simulations ``out of distribution'' from the perspective of a network having seen only Fiducial galaxies during training, which explains the increased errors in reconstruction.

This Figure also shows that $Z_*$ breaks the degeneracies between these two parameters, as the parameters have an opposite effect on this feature. While this is observed in both \texttt{IllustrisTNG} and \texttt{ASTRID}, the asymmetry in the effect of \om{}--and thus the discrepancies with the effect of $A_{SN1}$-- is more pronounced in \texttt{IllustrisTNG}. 

As expected, the average displacement vector for $A_{SN2}$ is very different in \texttt{ASTRID}, as this specific parameter is implemented in a different fashion. Given the magnitude of the effect of $A_{SN2}$ in \texttt{ASTRID} --and putting it in perspective with the magnitude of the effect of \om-- it can explain why, in \cite{Chawak2024ApJ}, the prediction precision and accuracy for $A_{SN2}$ saturates at 4 galaxies for \texttt{ASTRID} (see Fig 3 in \cite{Chawak2024ApJ}): a change in $A_{SN2}$ induces a strong enough difference and separation in the resulting distribution that it can be characterized with only 4 randomly picked galaxies. 
  
In a general sense, the effect of increasing \om{} seems consistent between the two suites in most of the features --with the caveat, however, that the suites have a different fiducial distribution to start with, as we show below. We observe stronger discrepancies between the two suites when decreasing \om{}, especially for $M_*, M_{BH}$ and $R_*$.

Broadly speaking, the \om{} trends seen in the top panels of Figure~\ref{fig:OT} make sense in terms of the physical model of cosmology, in which mergers of dark matter substructures dominate structure formation at small scales.
For example, as the matter density \om{} increases, galaxies are generally expected to have larger total masses, larger characteristic velocities, larger velocity dispersions, and larger peculiar motions. This is because as \om{} increases, the dynamical effect of gravity increases, infall timescales are shorter; gravitational collapses are larger and more complete at the present day. Gravitational perturbations from nearby large-scale structure are also larger.
The gas-phase and stellar metallicity trends with \om{} also make sense in this simple picture since the retention and growth of metals, especially at early times, is expected to depend on the gravitational binding energies of the stellar populations producing the supernovae. Gravitational binding energies generally increase with \om{}.

The second row of panels in Figure~\ref{fig:OT} raises a question: Why are the $\sigma_8$ trends smaller in amplitude than the \om{} trends?
After all, both \om{} and $\sigma_8$ enter in the growth of structure, but with different functional relationships to the nonlinear growth of structure:
$\sigma_8$ controls the initial conditions or ``starting point'' of the matter fluctuations, while \om{} controls how quickly they grow from that starting point.
In the linear regime, both of these linearly relate to the amplitude of the growth factor at the present day. 
However, in the nonlinear regime (and the galaxies in question are very strongly in the nonlinear regime), increasing \om{} shortens all orbital and infall timescales, hastening mergers and growth, whereas the initial conditions (controlled by $\sigma_8$) are increasingly forgotten, dynamically.
Thus although $\sigma_8$ and \om{} are expected to have qualitatively similar effects on large-scale structure, at galaxy scales, we expect \om{} effects to be amplified relative to $\sigma_8$ effects.
This expectation is borne out in Figure~\ref{fig:OT}.

Despite $\sigma_8$ effect being less pronounced than for \om, it is non-negligible (and also seen in Figure \ref{fig:accuracyClassif_1P}), indicating that the global distribution is indeed changing: this is thus consistent with the results observed in \cite{Chawak2024ApJ}, where increasing the number of galaxies used (hence getting more information about the full distribution) steadily reduced the error in prediction for that parameter. However, that work did not investigate the robustness across simulations for $\sigma_8$, and Figure \ref{fig:OT} might indicate some differences in the effect of $\sigma_8$ between \texttt{ASTRID} and \texttt{IllustrisTNG}, for instance, on the star formation rate (SFR).

\begin{figure*}
    \centering
\includegraphics[width=\linewidth]{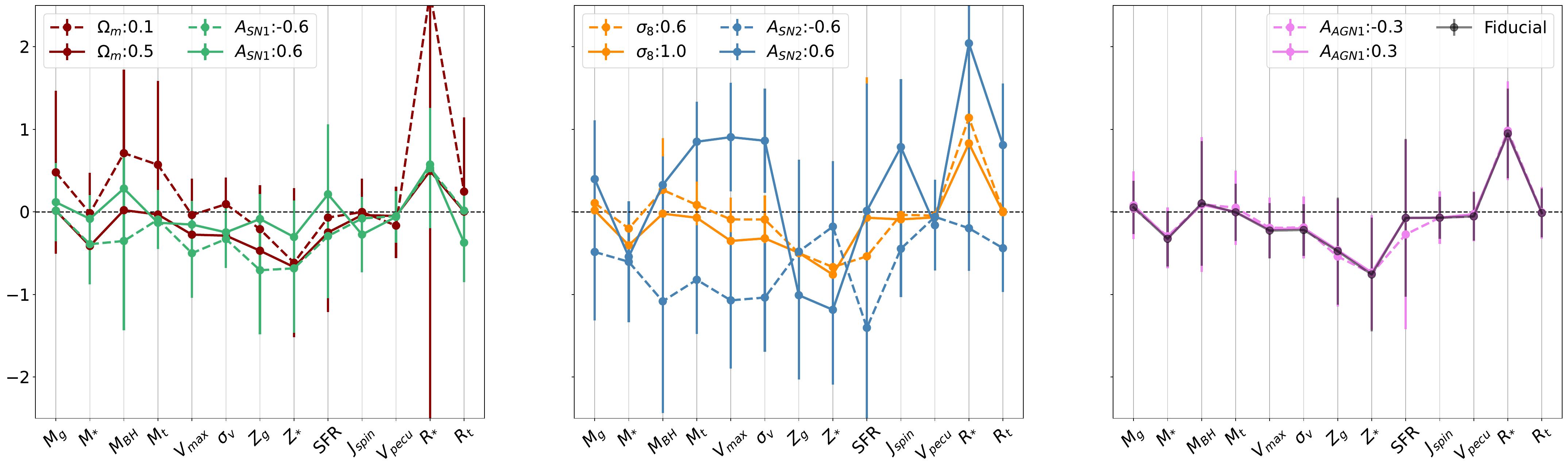}
    \caption{Average Displacement Vector (ADV) computed using Optimal Transport, between simulations from \texttt{IllustrisTNG} and \texttt{ASTRID} from the 1P suite (Varying only one parameter at a time). Color and line-style indicate the parameter varied (all others being fixed to fiducial values), e.g. dark-red solid line shows the displacement between \texttt{IllustrisTNG} and \texttt{ASTRID} galaxies for \om$=0.5$. Each simulation dataset was standardized beforehand with respect to the fiducial simulation from \texttt{IllustrisTNG}; the y-axis shows the ADV in each dimension in the normalized space.}%
    \label{fig:OT_illuTOastrid}
\end{figure*}

It is more difficult to develop clear physical intuitions for the differences in sensitivities to (and effect of) cosmological parameters between \texttt{ASTRID} and \texttt{IllustrisTNG}. Moreover, this analysis does not account for existing differences in the starting fiducial distribution, for instance, between the two suites. To investigate this aspect, we visualize the Average Displacement Vectors in Figure \ref{fig:OT_illuTOastrid}, computed in the full (13) feature space, between \texttt{IllustrisTNG} and \texttt{ASTRID} simulations with identical parameters. We standardize the simulations with respect to the fiducial \texttt{IllustrisTNG}. The color and line style indicate the parametrization of the simulations considered, where we either compare the two fiducial simulations (black line, right panel), or simulations varying only one parameter. A positive  (resp. negative) value for a feature $x$ indicates that the mean of the distribution for $x$ is larger (resp. smaller) in the ``twin'' simulation from \texttt{ASTRID}. 

When comparing the fiducial simulations, we see that the two suites seem to be in good agreement for most mass and velocity features, in terms of the mean at least. The metallicities ($Z_*, Z_g$) and $R_*$, on the other hand, exhibit stronger discrepancies. It is however relevant to point out that an agreement in this mean (i.e., a value close to 0 in Figure \ref{fig:OT_illuTOastrid}) can hide different underlying distributions: for instance, we observe for $M_{BH}$ that \texttt{ASTRID} tends to create more galaxies with $M_{BH}=0$, but also more massive black holes when they do exist (see Figure \ref{fig:hist_MBH} in Appendix \ref{sec:OTviz}). This gets reflected on the error-bars for that feature. In the case of $M_{BH}$, this could be explained by the fact that \texttt{ASTRID} seeds black holes in haloes with a mass an order of magnitude smaller compared to \texttt{IllustrisTNG}; these black holes would thus have more time to grow until $z=0$. Additionally, the rules implemented in the two simulations for black hole accretion and feedback are very different and surely play a role in explaining those discrepancies. 

The displacement vectors between simulations varying $A_{SN2}$ (middle panel, blue solid and dashed lines) indicate strong discrepancies between the distributions: this is expected since this parameter is implemented in a significantly different way in \texttt{ASTRID}.
From the left panel, we can observe that the displacement computed between the simulations with \om{} $=0.5$ (dark red solid line) appears similar to the fiducial one: we interpret this as an indication that this parameter induces similar changes in both suite --thus keeping the difference between the simulations similar, while for \om{} $=0.1$, the simulations diverge more (see $M_{BH}, M_t, R_*$). This hypothesis is consistent with Figure \ref{fig:OT}, which showed that lowering \om{} had a more significant effect on those features in \texttt{IllustrisTNG}.

We note that those discrepancies make the simulations {\em with the same parameters} relatively easy to distinguish: using a similar setup as in Section \ref{sec:paramimpact}, we obtain $\sim 85\%$ accuracy in classifying galaxies from the two simulations when using only $M_g, M_*, \sigma_v$ and $Z_*$. This means that those simulations, despite having the same cosmological and astrophysical parameters, produce distributions that populate different parts of the feature space. Those discrepancies also explain why the ability to transfer between suites is dramatically impacted when using several galaxies in \cite{Chawak2024ApJ}, as providing more than one galaxy gives additional information on the overall distribution, which an ML model will pick up on.

\section{Discussion}
\label{sec:discu}

We can summarize our findings as follow:

\begin{itemize}
    \item In the suites of simulations that we have investigated, we confirm that \om{} significantly impacts the distribution of a subset of galaxy properties in specific directions: that is, for a ``small'' change in \om, there is a large enough shift of the resulting distribution (compared to its global variance as well) so that the overlap between two simulations with different \om{} values is relatively small. This was expected from the results of \cite{onegalcosmo} since if \om{} had little or no effect, it would not be predictable, even less so from a single galaxy.
    \item There are almost no other parameters ---cosmological or astrophysical--- in the suite that move the distribution of galaxy properties either in a similar way or in opposite ways to \om{} with the same magnitude: that is, there is no mechanism to create degeneracies or confounders in terms of distributions. Some parameters have barely any impact --which also explains why these parameters were ``not predictable'' in \cite{onegalcosmo} and other works. While $A_{SN1}$ does create some degeneracies in part of the feature space, other features allow breaking those degeneracies. We also note that creating efficient degeneracies for 13 features would require either many parameters with strong effects or a parameter that has a roughly similar effect as \om{} in the full feature space. 
    
    This aspect, combined with the previous point, explains and confirms the predictability of \om{} from a single (or few) galaxies in the context of the considered suites of simulations: \om{} has a strong effect, and no mechanism obfuscates it, making marginalizing over other parameters trivial.
    
    \item Estimations of the effective impact of \om{} on the feature space provide a way to build hypotheses regarding the underlying physical mechanisms. The trend seems reasonable with what we could expect from a cosmological standpoint. However, \texttt{ASTRID} and \texttt{IllustrisTNG} show some discrepancies in the effect of \om{}(and $\sigma_8$) for a subset of  properties. A significant difference in the effects of cosmological parameters --induced here likely by hydro-code or subgrid physics-- could become problematic for further cosmological analyses using those properties. More work is required to fully quantify this aspect, as the suites also show dissimilarities in the distributions of galaxy properties for identical cosmological and astrophysical parametrization, thus making the comparison more difficult. This is also partially expected but is perhaps more concerning for the fiducial case. 
\end{itemize}

We note that using more features to predict \om{} should only improve accuracy (in the limit of the model used to handle uninformative or noisy features), as more features will only help further break any residual degeneracies.
Additionally, the ``predictability'' of \om{} can hold even when integrating more (astrophysical or cosmological) parameters if those parameters do not impact the relevant subset of features significantly enough (i.e. introducing enough variance or confounders). 

However, the crux of the problem --as exhibited in practice in \cite{casco2023}-- lies indeed in the differences between the distributions of galaxy properties across different suites and how the parameters affect those distributions from one suite to the next. We currently do not have an explicit quantitative model of the effect of \om{} all the way to galaxy properties level. 
On the astrophysics side, many of the processes explored --and their associated parameters-- are evaluated and calibrated using observations combined with strong priors on cosmology. Sadly, we only get to observe a single Universe. Thus, bringing these results to actual data and deriving cosmological parameters from them would be a significant challenge.

On the other hand, this work suggests potential avenues and tools of analysis to deepen our understanding of the interplay between the hydro-simulations and the effects of cosmological parameters on less investigated statistics and observables such as galaxy properties: this aspect is also crucial for astrophysical prospects relying on simulations, since we actually have --at least in the context of $\Lambda$CDM-- relatively tight constraints on \om{} \citep{planck2018, cosmo2020boss}. Those existing constraints motivate the investigation of simulations with a tighter range on \om. These would also allow studying the effect of \om{} effect at smaller variation scales.

One could be tempted to address the problem of differing simulations by identifying features that behave consistently across simulations, under the parameter of interest, in the hope of finding features that will transfer well to real data as well. However, such approaches --aiming at building methods ``robust'' across simulation models-- rely on the following assumptions in order to be beneficial for real-data applications: (i)``reality'' has to be covered by (or close enough to) the simulations,(ii) the simulations accurately represent the assumption that went into them, (those two being usual assumptions of simulation-based inference), but additionally (iii) all the suites of simulations (or models) are considered to be somewhat equally or comparatively ``right'' or close to the ``truth''. Those assumptions are quite strong and, unfortunately, hard to meet (or even evaluate) in most cases.

In the future, it would be interesting to refine the analyses using Optimal Transport on restricted sets of similar galaxies (that is, selecting a specific population of galaxies beforehand, e.g. in terms of mass range etc), to potentially eliminate the signal coming from the discrepancies in original distributions between suites. Interestingly, all simulations in the \texttt{CAMELS LH} suites share initial conditions. This means that we could potentially match galaxies across different simulations and investigate directly how they change individually instead of using Optimal Transport. \texttt{CAMELS} also provide the CV set --which was leveraged in \cite{lue2025iobonegal}: simulations varying the initial conditions but with fixed fiducial parametrization. These sets would allow investigating the impact of cosmic variance on the galaxies' properties distributions, and putting it in perspective with the parameters' effects. Finally, conducting a more theoretically grounded work to derive --or validate-- a model of \om's effect on galaxies' properties would be essential to address some of the questions that \cite{onegalcosmo} and other works brought up.

\begin{acknowledgments}
We thank Mike Blanton, Shy Genel, Chris Lovell, Amanda Lue, Chirag Modi, Oliver Philcox for useful insights and comments.
The Flatiron Institute is a division of the Simons Foundation. GC acknowledges support by the European Union's Horizon Europe research and innovation programme under the Marie Skłodowska-Curie Postdoctoral Fellowship Programme, SMASH co-funded under the grant agreement No. 101081355. The operation (SMASH project) is co-funded by the Republic of Slovenia and the European Union from the European Regional Development Fund. The work of FVN is supported by the Simons Foundation. RT acknowledges co-funding from Next Generation EU, in the context of the National Recovery and Resilience Plan, Investment PE1 – Project FAIR ``Future Artificial Intelligence Research''. This resource was co-financed by the Next Generation EU [DM 1555 del 11.10.22]. RT is partially supported by the Fondazione ICSC, Spoke 3 ``Astrophysics and Cosmos Observations'', Piano Nazionale di Ripresa e Resilienza Project ID CN00000013 ``Italian Research Center on High-Performance Computing, Big Data and Quantum Computing'' funded by MUR Missione 4 Componente 2 Investimento 1.4: Potenziamento strutture di ricerca e creazione di ``campioni nazionali di R\&S (M4C2-19 )'' - Next Generation EU (NGEU). 

\end{acknowledgments}

\appendix

\section{Impact of the Latin Hypercube boundaries}
\label{sec:boundaries}

Having established that indeed some of the parameters, and in particular $\Omega_{\rm m}$ and $A_{SN1}$, lead to detectable shifts in the distribution of features in different simulations (at least when varied individually), we want to investigate the impact of the edges and span of the Latin Hypercube on the predictability of cosmological parameters, in particular ${\Omega}_m$. 

It can be argued that if the ability to predict $\Omega_{\rm m}$ accurately from the features $\Sfeat{i}{L}$ of a single galaxy $i$ is independent or robust to the choices on the astrophysical processes --``marginalizing'' over them-- then it should be robust specifically to changes or shifts in the ranges of priors assumed for the other parameters in the training suite.
Since we cannot check for robustness when extending the parameters' ranges (as those simulations do not exist and would be computationally expensive to obtain), we resort to the opposite test. We restrict or shift the LH range to ascertain whether the prediction task becomes easier (or harder), which might indicate that a reduced/changed prior range has an impact on the classification accuracy. Conversely, this would lead to suspect that the accuracy obtained with the full LH might decrease if one had trained on a larger prior range. 

We define new ``sub Latin Hypercubes'' (sub-LH) from the original one. For each simulation in this new cube, we select 50 galaxies for training and 50 galaxies for testing. Note that this means that we do not split simulations in a train and test as in \cite{onegalcosmo}, which makes our prediction task slightly easier as we do not need to interpolate for unseen simulations but only unseen galaxies. Also, the number of training and testing examples is a function of the volume of the sub-LH and we must be careful when comparing performance metrics between sub-LH of different volumes. We then fit a Gradient Boosted Tree model with 40 estimators and the default settings of \texttt{xgboost} \citep{xgboost}, using either all the features or a subset of them, to predict \om. We evaluate the Median Absolute Error (MAE) in prediction on the held-out test galaxies. We repeat this operation 50 times by resampling the training and testing set to estimate the mean and variance of the MAE in a given sub-LH.

We design two different ways of creating sub-LHs, but always varying the range of a single parameter at a time:
\begin{enumerate}
    \item \textbf{Shifting sub-LH}: we fix the volume (i.e. the length of the range along one parameter axis) to one-third of the original range, and we create sub-LHs by moving along the original range of the parameter in steps of size 10\% of the original range (denoted in the Figures as the ``decile starting point"). For instance, for parameter $A_{SN1}$, of original range (in log-space, to keep the distribution uniform) $[-0.6, 0.6]$, we create sub-LHs spanning $[-0.6, -0.2]$ (decile 0), $[-0.48, -0.08]$ (decile 1), etc.
    
    In this case, all datasets span the same sub-volume of the original LH but cover different regions of a given parameter. Thus, an increase (decrease) in performance means that one of the regions features less (more) variance than the other. On the other hand, if the performance is similar, this could be due to either (i) the parameter introducing comparable variability across the different ranges (e.g. a linear impact on the features as a function of the parameter), or (ii) the parameter being associated with very small (or negligible) variability.
    \item \textbf{Shrinking sub-LH}: centered on the mean, we reduce the parameter's range symmetrically. For instance, for parameter $A_{SN1}$, we create sub-LHs spanning $[-0.6, 0.6]$ (100\%), $[-0.51, 0.51]$ (85\%), $[-0.42, 0.42]$ (70\%), etc. We express the resulting sub-LHs in terms of the percentage of the original LH preserved.
    
    By decreasing the parameter range, we either maintain or decrease the variability of the data. If the variability remains similar, the performance could either decrease (as we have less examples to train on) or stay the same. If the variability decreases, the performance should improve.
\end{enumerate}

\begin{figure*}
    \centering
    \includegraphics[width=\linewidth]{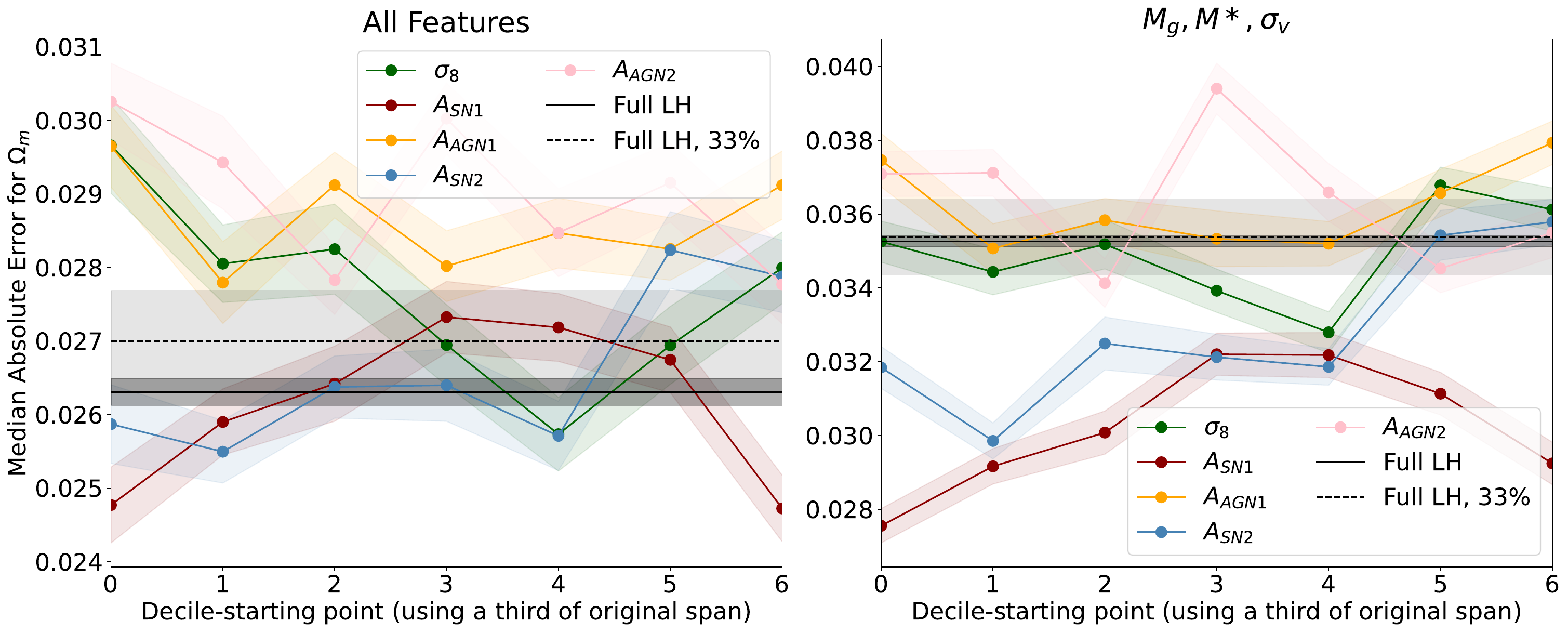}
    \caption{Mean and variance of the Median Absolute Error (MAE) when training and predicting \om{} with different sub-LH following the \textbf{shifting sub-LH} protocol: each sub-LH spans one-third of the original parameter range (indicated by the color), and moves along the parameter values by step (indicated by ``decile starting point'') amounting to 10\% of the parameter's range. Left panel shows results using all the features. Right panel shows results using only $M_g, M_*,$ and $\sigma_v$. The baseline using simulations from the whole LH is shown in solid black, while dashed black shows the original LH but with simulations down-sampled to a third. Values under this line indicate an improvement in performance in predicting \om{}.}
    \label{fig:shiftLH}
\end{figure*}

\paragraph{Shifting sub-LH:}
Figure \ref{fig:shiftLH} shows the mean and variance of the Median Absolute Error (MAE) for predicting \om{} when shifting the LH along each of the remaining 5 parameters' direction, using all features on the left panel, and only three features ($M_g, M*$ and $\sigma_v$) on the right panel. We show in solid black the baseline using the full LH, and in dashed black a baseline using the full LH downsampled to a third of the simulations (i.e. a similar sized dataset ---in terms of the number of training and testing examples but not volume--- to the other experiments of the figure), for comparison. We note that results for nearby deciles (for individual parameters) are correlated, as there is an overlap in the shifted sub-LHs. 

Using only three features (right panel), we see that restricting the span of $A_{SN1}$ to the sub-LH (dark red) improves accuracy for all decile choices when compared with the accuracy obtained from the full LH (black baseline). Values of $A_{SN1}$ near the boundaries of the LH (decile starting points 0 and 6) lead to even better accuracy, which we interpret as extreme values of $A_{SN1}$ introducing less variability than middle values. We also see this effect for $A_{SN2}$, where lower starting deciles show an improvement in MAE. 

This can be interpreted as (i) the full range of $A_{SN1}$ and $A_{SN2}$ introduces more variance (or degeneracies) than a restricted range, (ii) higher values of $A_{SN2}$, and middle values of $A_{SN1}$, create more confounders in the data; this is consistent with the analysis in Section \ref{sec:degeneracies}. Shifting the ranges of the other parameters does not seem to impact predictability, and even with much fewer data, models can train and generalize to the same level as the full-LH (compare yellow and pink lines ($A_{AGN1}$ and $A_{AGN2}$) with solid black baseline on right panel of Figure \ref{fig:shiftLH}). This is  also in line with the observations made in Section \ref{sec:paramimpact} showing a lack of signal from those parameters.

When using all features (left panel), we see a similar trend for $A_{SN1}$, albeit less pronounced, and the effect of $A_{SN2}$ mostly disappears. The predictive performance also decreases for some of the parameters: this could be explained by having the same amount of variability, but only a third of data to train in a higher dimensional feature space, which could lead to either under-fitting or over-fitting, both of which would decrease predictivity. However, the models underperform compared to a baseline using a third of the data as well.

\begin{figure*}
    \centering
    \includegraphics[width=\linewidth]{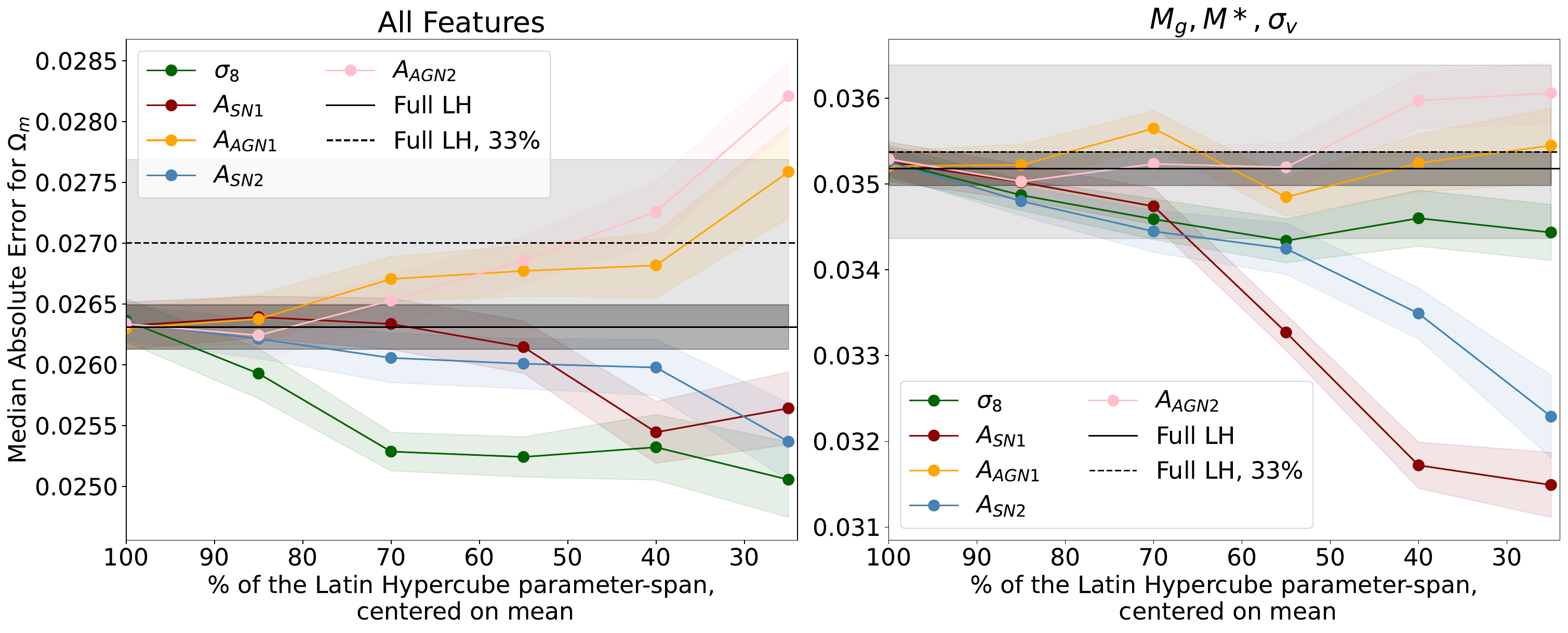}
    \caption{Mean and variance of the Median Absolute Error (MAE) when training and predicting \om{} with different sub-LH following the \textbf{shrinking sub-LH} protocol, where the x-axis indicates the amount of the parameter (indicated by the color) range kept to create the sub-LH (centred on the mean of the parameter). Left panel shows results using all the available features. Right panel shows results using only $M_g, M_*,$ and $\sigma_v$. The baseline using the entire LH is shown in solid black, while the dashed black line shows the performance using the full LH downsample to a third of the simulations.}
    \label{fig:shrinkLH}
\end{figure*}

\paragraph{Shrinking sub-LH:}
We plot in Figure \ref{fig:shrinkLH} the mean and variance of the MAE for \om{} when shrinking the LH for each of the remaining five parameters. We show the performance using the full LH as a reference baseline in solid black, and using the full LH but only a third of simulations in dashed-black. Again, the left panel shows results using all features, and the right panel shows results using only $M_g, M*$ and $\sigma_v$. In terms of which parameters' training range has an appreciable influence on predictive performance, we observe similar trends to the shifting sub-LH experiments: reducing the range (around the mean) of $A_{SN1}$ and $A_{SN2}$ improves performance in the 3-feature cases, and to a lesser extent this is also observed for $\sigma_8$. This could be interpreted as ``shaving'' the edges of $A_{SN1}$ and $A_{SN2}$ removes potential degeneracies --at least in the 3-feature space--, thus improving performances. In the full feature space, this trend mostly disappears for $A_{SN2}$. The trend persists, albeit to a much lesser extent (actual MAE change is small), for $A_{SN1}$, meaning that $A_{SN1}$ still create some degeneracies or variance, but it is better mitigated or less significant with the full set of features. It seems the introduction of degeneracies ``plateaus'' at $\sim$ 70\% (i.e. $ -0.421 < A_{SN1} < 0.421$), which is coherent with Figure \ref{fig:shiftLH} where extreme values of $A_{SN1}$ seem to introduce less variability (better performance in Fig \ref{fig:shiftLH}).  Removing regions with low variability in themselves might not improve performance so much; however now we learn also that those parameters regions are not creating degeneracies or confounders for other regions in the full cube. However, more central values of $A_{SN1}$ seem to introduce comparatively more degeneracies. 

Interestingly, the trend remains also for $\sigma_8$, where we observe a change in trend on the performance (degrading for LH $>$ 70\%, i.e. for $\sigma_8 < 0.66$ and/or $\sigma_8 > 0.94$) which can mean that these $\sigma_8$'s regions start introducing degeneracies or variance. 

Comparatively, we see that the performance when shrinking the LH along $A_{AGN1}$ and $A_{AGN2}$ degrades to a similar performance when using the full LH with 30\% of the data (dashed black line), which is coherent with those parameters not impacting the variability of the data.

\section{Optimal Transport Visualization}
\label{sec:OTviz}

\begin{figure}
    \centering
    \includegraphics[width=0.9\linewidth]{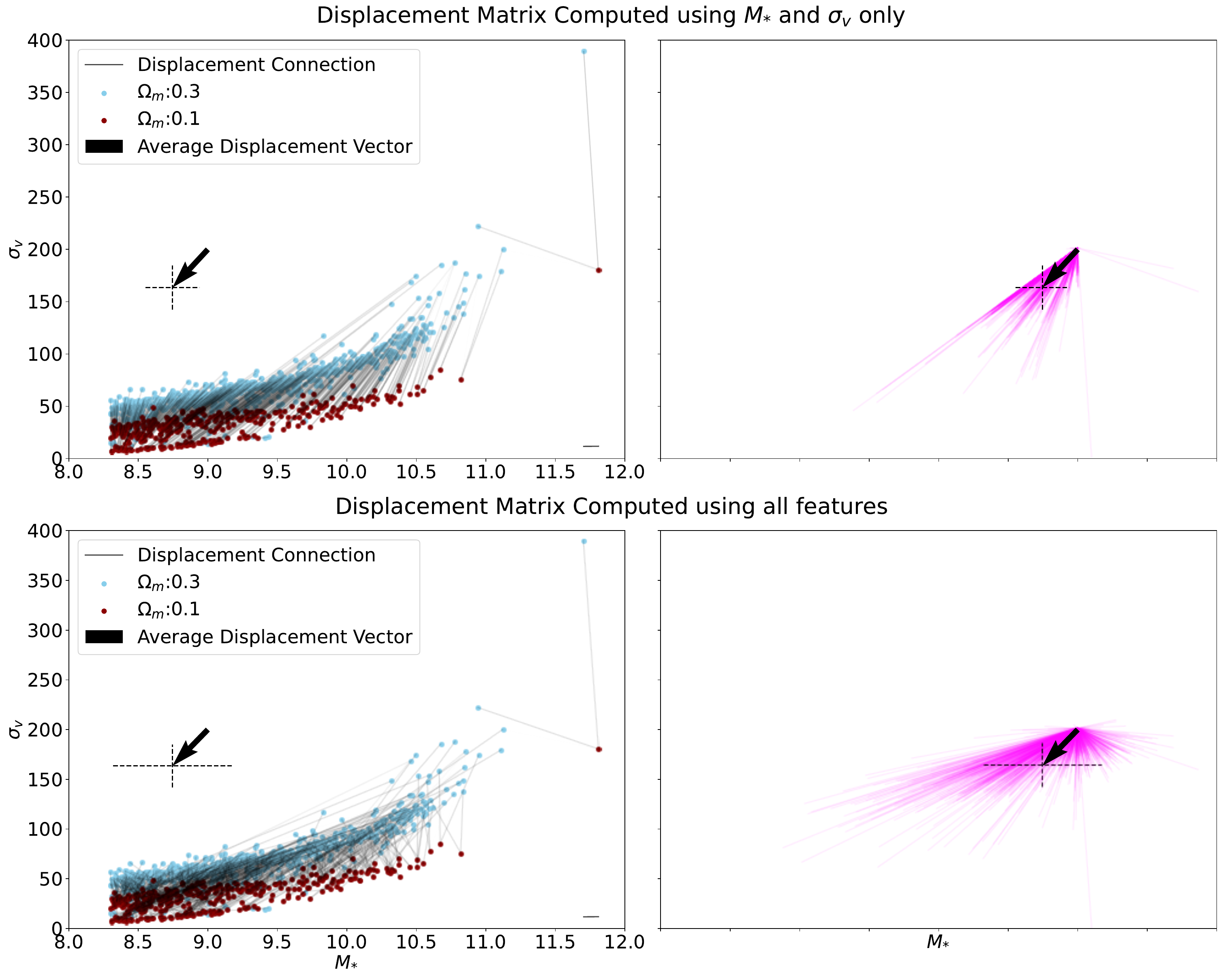}
    \caption{Left: Galaxies from \texttt{IllustrisTNG}-1P simulations (fiducial in blue, and with \om$ = 0.1$ in red), displayed in $M_*/\sigma_v$ space, where the Optimal Transportation Matrix $\boldsymbol \gamma$ is depicted as black lines between two galaxies if the value of the matrix for that pair is non-null. The resulting average displacement vector computed is displayed as the black arrow on the left quadrant of the plots. Right: the same connections (black lines on the left plot) are depicted as pink lines, starting from the same origin, with the average displacement as a black arrow. The difference between the top and bottom figures stems from the feature space in which the Transportation matrix (that is, the distance matrix it relies on) is computed: top row is computed in the 2D feature space depicted. Bottom row uses the full 13 features space. This illustrates how the average vector is the same in both cases, but the variance in the displacement is greater in the full feature space --as the neighborhood of a galaxy is impacted.}
    \label{fig:OTIllu2D}
\end{figure}

Figure \ref{fig:OTIllu2D} illustrates the optimal transportation matrix and the information we derive from it computed between two simulations: the first row compute the OT calculation using only two galaxy properties, $M_*$ and $\sigma_v$; the second row uses the entire 13 features. We depict in blue galaxies from \texttt{IllustrisTNG}-1P fiducial, and in dark red galaxies from \texttt{IllustrisTNG}-1P with \om$=0.1$. The black lines connecting the blue and red dots indicate that there is a connection between those points in the transportation matrix (i.e. $\boldsymbol\gamma_{i,j}>0$). The black arrow indicates the average displacement vector, and the dashed lines show the standard deviation computed from $\boldsymbol \gamma$ as described in Section \ref{sec:astrid}. The right-side panels show similar information but plot all the connecting vectors with the same starting points for illustrative purposes.

We can see that while the average displacement vector is the same in the first and second row, the actual transportation matrix $\gamma$ is different, thus impacting the standard deviation of the resulting average displacement vector. 

\begin{figure}
    \centering    \includegraphics[width=0.5\linewidth]{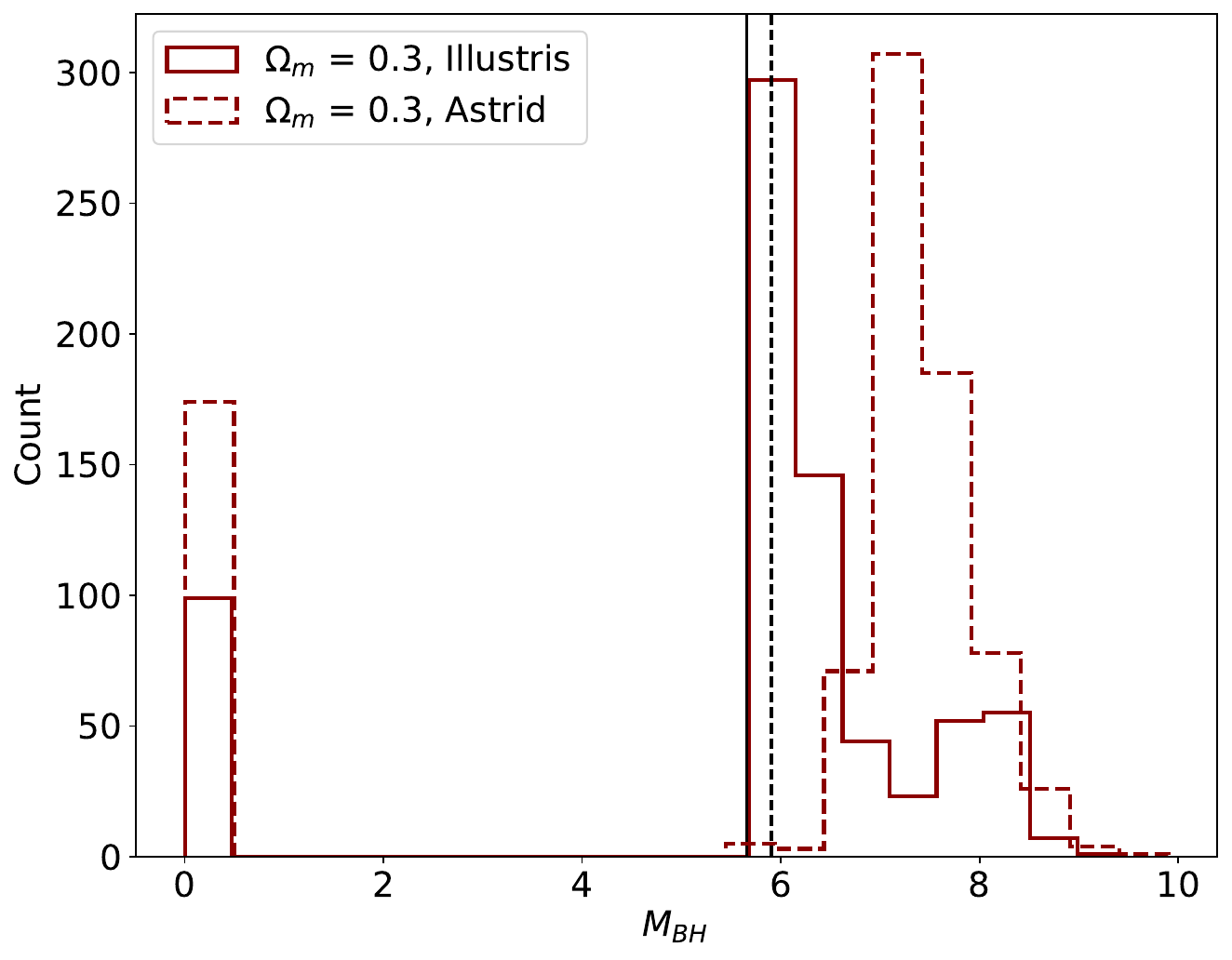}
    \caption{Histogram count of galaxies' $M_{BH}$ in the 1P fiducial simulation of \texttt{IllustrisTNG}  (solid line) and \texttt{ASTRID} (dashed). Their respective means are indicated as vertical black lines with consistent linestyle. }
    \label{fig:hist_MBH}
\end{figure}

Figure \ref{fig:hist_MBH} shows the histogram of the galaxies' $M_{BH}$ in the 1P fiducial simulation of \texttt{IllustrisTNG}  (solid line) and \texttt{ASTRID} (dashed), as well as their respective means with consistent line styles. This illustrates the limits of our analysis in Section \ref{sec:astrid}, where we use the mean as a summary statistic, as we do not account for this type of bimodality leading to two very different distributions with a similar mean. 

\bibliography{biblio}{}
\bibliographystyle{aasjournal}

\end{document}